\documentclass[sn-mathphys,Numbered]{sn-jnl}
\usepackage{graphicx}
\usepackage{multirow}
\usepackage{amsmath,amssymb,amsfonts}
\usepackage{amsthm}
\usepackage{mathrsfs}
\usepackage[title]{appendix}
\usepackage{xcolor}
\usepackage{textcomp}
\usepackage{manyfoot}
\usepackage{booktabs}
\usepackage{algorithm}
\usepackage{algorithmicx}
\usepackage{algpseudocode}
\usepackage{listings}
\usepackage{soul}
\usepackage{array}
\usepackage{comment}

\begin{document}

\title{Accelerating the drive towards energy-efficient generative AI with quantum computing algorithms}

\author[1,2]{\fnm{Frederik F.} \sur{Flöther}}

\author[1]{\fnm{Jan} \sur{Mikolon}}

\author[3]{\fnm{Maria} \sur{Longobardi}}

\affil[1]{\orgname{QuantumBasel}, \orgaddress{\street{Schorenweg 44b}, \city{Arlesheim}, \postcode{4144}, \country{Switzerland}}}

\affil[2]{\orgname{Center for Quantum Computing and Quantum Coherence (QC2), University of Basel}, \orgaddress{\street{Petersplatz 1}, \city{Basel}, \postcode{4001}, \country{Switzerland}}}

\affil[3]{\orgname{NCCR SPIN, University of Basel}, \orgaddress{\street{Klingelbergstrasse 82}, \city{Basel}, \postcode{4056}, \country{Switzerland}}}

\maketitle
\begin{abstract}

\textbf{Abstract.} Research and usage of artificial intelligence, particularly generative and large language models, have rapidly progressed over the last years. This has, however, given rise to issues due to high energy consumption. While quantum computing is not (yet) mainstream, its intersection with machine learning is especially promising, and the technology could alleviate some of these energy challenges. In this perspective article, we break down the lifecycle stages of large language models and discuss relevant enhancements based on quantum algorithms that may aid energy efficiency and sustainability, including industry application examples and open research problems.

\end{abstract}
\section{The need for quantum-enhanced efficient generative AI}

The last years have seen an explosion in the development and adoption of artificial intelligence (AI) models and algorithms. In particular, generative AI techniques~\cite{sengar2024generative} and large language models (LLMs)~\cite{naveed2023comprehensive} have shown astonishing improvements. At the same time, there are a number of risks with this new generation of models~\cite{bommasani2021opportunities, wach2023dark}. The energy consumption and environmental impact~\cite{george2023environmental}, for both the training and inference stage~\cite{desislavov2023trends}, are among the most important challenges. For instance, the OpenAI o3 and DeepSeek-R1 models consume over 33Wh per long prompt~\cite{jegham2025hungry}.

Quantum computing is at an earlier stage than AI; it has not yet experienced commercial breakthroughs. The technology represents a fundamentally different approach to processing information, using quantum bits (qubits) instead of bits. Through the clever exploitation of quantum mechanical effects such as entanglement, interference, and superposition, novel algorithms become possible which allow significant, in some cases exponential, improvements compared with classical techniques~\cite{gill2022quantum}. It is, in fact, the only known computational model which enables such exponential speedups~\cite{horowitz2019quantum}. The potential benefits go beyond speed; quantum computers might also allow AI models to be developed and calculations to be performed with greater accuracy, higher energy efficiency, and lower input data requirements (in terms of quality and quantity). In fact, there is also a symbiosis as AI, including LLMs, is being explored to accelerate the development of quantum technology~\cite{krenn2023artificial,liang2023unleashing,alexeev2024artificial} while quantum computing is being researched for a myriad of machine learning tasks~\cite{peral2024systematic}.

Despite quantum computing's considerable potential, its energy efficiency is still being debated~\cite{chen2023quantum}. The controlled laboratory environment required by leading quantum computing architectures also generates significant power demands, which, depending on the exact technology, may have an order of magnitude of tens of kW~\cite{desdentado2024exploring}. Comparing that with the MW power that classical supercomputers often require suggests that quantum computers can be competitive in terms of energy consumption-related quantum advantages.

One of the challenges of applying quantum algorithms to problems involving classical data has been the ``input problem", which refers to the fact that efficiently loading large volumes of classical data into today's quantum computers remains difficult~\cite{duan2024compact}. While research is trying to address this issue, involving, for instance, quantum random access memories (QRAMs)~\cite{phalak2023quantum} and coresets~\cite{harrow2020small, yogendran2024big}, loading larger volumes of classical data today is not feasible or might actually erase any quantum advantages.

One reason why modern generative AI requires so much energy is the size of the models. Cutting-edge models typically have billions of parameters, which enable them to achieve optimal performance~\cite{gholami2024generative}. There are research efforts underway to make AI models smaller~\cite{jovanovic2024compacting}. Another promising avenue that has emerged with regard to creating more efficient architectures is based on a class of brain-inspired continuous-time recurrent neural networks called liquid neural networks~\cite{chahine2023robust}. As AI models become more compact, they are becoming more and more amenable to integration with quantum algorithms due to the aforementioned ``input problem", opening up the possibility of further training efficiency gains.

Given the proliferation and increasing energy consumption of AI, the potential of quantum computing to drive efficiency gains in this space is of great interest. The present discussion focuses on quantum computing algorithms that can significantly increase the efficiency of certain key tasks in the lifecycle of LLMs. Clearly, there are other ways in which quantum computing techniques may address AI energy consumption challenges, which could be in the inference stage or an entirely different application space, such as the optimization of renewable energy integration in AI-driven data centers~\cite{veeramachaneni2025optimizing}.

\section{Applications of quantum algorithms in generative AI development}

Table \ref{table:algos} summarizes a selection of promising quantum algorithms and techniques which lend themselves to near-term and long-term enhancements of classical AI model training/inference efficiency.

\begin{table}
    \centering
    \renewcommand{\arraystretch}{1.5}
    \begin{tabular}{m{2.7cm} m{2.7cm} m{2.7cm} m{2.7cm} m{2.7cm}}
       LLM lifecycle stage & Classical approach & Possible quantum enhancement & Sustainability rationale & Time scale, Potential\\
    \hline
    \hline
         Data collection and curation & Massive web scraping, distributed data deduplication and filtering & Quantum-assisted clustering/deduplication (via advanced sampling)~\cite{poggiali2024quantum} & Reduced redundant data lowers overall data processing/storage costs & Medium-term, low\\
         Preprocessing and encoding & Text tokenization (byte-pair encoding, WordPiece) & Compact data-loading circuits (e.g. QRAM~\cite{jaques2023qram}, amplitude encoding~\cite{weigold2021encoding}) & Potentially fewer large-scale CPU/GPU cycles used in repeated data transformations & Long-term, low\\
         Model initialization and architecture & Random weight initialization (Xavier, Kaiming), billion+ parameter models & Quantum hyperparameter search~\cite{wulff2024distributed}, hybrid quantum neural network layers~\cite{zhao2021review} & Smaller, more expressive models can lower energy consumption & Medium-term, high\\
         Training (core loop) & Stochastic gradient descent, Adam, large-scale distributed training, mixed-precision training & Quantum gradient methods~\cite{kerenidis2020quantum}, quantum natural gradient~\cite{stokes2020quantum}, quantum approximate optimization algorithm (QAOA)~\cite{zhou2023qaoa} & Fewer iterations/epochs lead to lower energy usage in high-performance computing (HPC) clusters & Medium-term, medium\\
         Training (fine-tuning and distillation) & Domain-specific fine-tuning, knowledge distillation, pruning & Quantum-assisted low-rank approximation~\cite{kong2025quantum}, quantum-based distillation and quantum reinforcement learning~\cite{meyer2022survey} for fine-tuning and knowledge distillation & Smaller distilled models reduce energy usage for both training and inference, yet can still achieve --- or even surpass --- the performance levels of larger counterparts & Near-term, high\\
         Inference and deployment & Quantization/model compression (e.g., INT8/FP16) for faster lower-memory inference & Quantum approximate optimization algorithm (QAOA) and quantum annealing~\cite{wang2026quantum} to identify which filters, neurons, or blocks in a network contribute least to accuracy and then pruning them & Faster inference time and hardware requirements while providing same performance levels as much larger models & Near-term, medium\\
         Maintenance and monitoring & Continuous monitoring, drift detection, logging of billions of requests & Quantum-accelerated anomaly detection~\cite{herr2021anomaly}, drift monitoring & Proactive retraining (done only when needed) lowers energy consumption & Medium-term, medium\\
         \hline
    \end{tabular}
    \caption{Overview of key LLM lifecycle stages, classical methods used therein, relevant quantum algorithms, and sustainability relevance as well as impact time scale and potential.}
    \label{table:algos}
\end{table}

The rightmost column gives a very rough assessment of the expected time scale when the quantum enhancements could generate value as well as the degree of their expected impact. For example, the lack of maturity of QRAM technology, and the relative ease with which classical techniques can handle preprocessing and encoding steps, suggests that this is an area where quantum computing may only generate value in the long term and the impact is likely low. Similarly, for the training (core loop) stage, the general need for big (classical) data processing makes it difficult for quantum computers to add value here in the near future. On the other hand, the training (fine-tuning and distillation) stage could see impactful quantum enhancements sooner, given that one can already conduct significant fine-tuning with on the order of 10--100 samples in certain cases.

Although the rapid advancements in quantum-assisted AI are very promising, real-world industry tests have only just begun. Still, they demonstrate the potential of quantum algorithms to enhance AI efficiency. The following application examples, aligned with Table \ref{table:algos}, provide practical evidence of how quantum computing may contribute to various stages of AI model development.

In the data collection and curation stage, quantum-enhanced clustering algorithms may support efficient data streamlining and storage. E.ON and Technical University of Munich demonstrated how a quantum \textit{k}-means algorithm can be adapted to real quantum hardware, clustering high-dimensional real-world German electricity grid data~\cite{diadamo2022practical}.

In the preprocessing and encoding stage, it is possible to leverage sophisticated data encoding in quantum algorithms. For instance, a group including Medical University of Vienna, Johannes Kepler University Linz, and Software Competence Center Hagenberg leveraged linear time quantum data encoding in the classification of clinical data~\cite{moradi2022clinical}.

In the model initialization and architecture stage, quantum algorithms may enable efficiency improvements through enhanced hyperparameter optimization. A consortium encompassing Lighthouse Disruptive Innovation Group, Vueling Airlines, University Ramon Llull, and MIT Media Lab employed a Fourier series method to represent the search space~\cite{consul2023quantum}. The data included flight no-show information as the dependent variable and features such as origin, destination, and time of flight. Variational quantum circuits were trained to select suitable hyperparameters. Furthermore, when novel architectures such as quantum neural networks are employed, both quantum~\cite{liao2024quantum} and classical~\cite{hu2022quantum} frameworks are being studied for tasks such as model training and compression.

In the training (core loop) stage, quantum algorithms may help tune the parameters of a classical neural network and overcome problems associated with backpropagation and gradient descent methods. Researchers from Politehnica University of Timișoara adapted Grover's algorithm in order to improve classical neural network weight optimization techniques, using the \textit{Digits} dataset from scikit-learn~\cite{jura2025quantum}.

In the training (fine-tuning and distillation) stage, researchers from IonQ introduced a hybrid quantum-classical deep learning architecture with the goal of improving token prediction for fine-tuned LLMs~\cite{kim2025quantum}. A group from Beijing University of Posts and Telecommunications explored a quantum knowledge distillation model for LLMs based on variational quantum circuits, using this approach to extract emotional information from text, detecting concealed sensitive information within media, and uncovering implicit themes in text~\cite{li2025quantum}.

In the inference and deployment stage, quantum algorithms could help further compress classical AI models, thus improving inference times. Large deep neural networks may often be significantly pruned while maintaining performance; the ``lottery ticket hypothesis" suggests that for certain architectures specific subnetworks may be found that reach comparable accuracies in a similar number of iterations as the original network~\cite{frankle2018lottery}. A Colorado State University researcher proposed several schemes, including variational quantum algorithms and quantum optimization methods, to enable quantum neuron selection with the goal of making this subnetwork discovery more efficient~\cite{whitaker2022quantum}.

Finally, in the maintenance and monitoring stage, quantum algorithms could improve the discovery of anomalies in order to better detect and prevent drift. A consortium including European Organization for Nuclear Research (CERN), IBM, and IRIS Analytics considered ensemble approaches where classical algorithms were combined with quantum feature selection in the context of fraud detection~\cite{grossi2022mixed}.

These industry efforts suggest that quantum computing could enhance energy efficiency throughout generative AI pipelines and help address other issues plaguing today's generative AI, such as hallucinations and low robustness. In fact, there are already early efforts that explore the implementation of transformer architectures and LLMs on quantum computers~\cite{liu2024towards,liao2024gpt,aizpurua2024quantum}.

\section{Outlook and open research challenges}

Although quantum computing holds considerable promise, most machine learning tasks will not see drastic benefits in the short term. Core processes like data preprocessing, feature engineering, and gradient-based optimization are already highly efficient on classical hardware involving central processing units (CPUs) / graphics processing units (GPUs) / tensor processing units (TPUs). Training deep learning models, especially transformers, relies on numerical methods that sometimes lack clear quantum speedups. Quantum computing is more likely to provide advantages in areas like quantum-enhanced kernel methods~\cite{yin2025experimental} or reinforcement learning. In reinforcement learning applications, quantum algorithms could enhance policy optimization, accelerate exploration through quantum sampling, and improve decision-making in complex environments. Quantum-inspired methods, as well as quantum annealing for solving multi-agent or high-dimensional problems, may offer further advantages over classical approaches. While still in its early stages, quantum-enhanced reinforcement learning could become another key area where quantum algorithms provide significant benefits.

Moreover, for quantum computing to realize its potential in enhancing AI efficiency, several fundamental challenges in hardware and architecture must be addressed. The realization of practical quantum advantages for AI requires significant improvements in areas such as the number of qubits, noise, memory efficiency, and system architecture, all of which currently limit the applicability of quantum algorithms in machine learning workflows. The limited number of qubits available in current systems is an important bottleneck in the adoption of quantum computing for AI, restricting, for instance, the number of features. In addition, tailored quantum hardware architectures for AI applications will also be critical for unlocking the full potential of quantum-enhanced machine learning. For example, in the same way that GPUs and other classical hardware is being optimized for AI training and inference tasks, quantum hardware and integrated quantum-classical architectures will need to be further fine-tuned to achieve optimal performance and energy efficiency. Early signs of that happening are already visible, given the increasing number of application-specific quantum algorithm tests as well as research around connecting multiple types of (different-qubit-modality) quantum systems together~\cite{su2023simple}.

\section{Acknowledgments}

This work was supported as part of NCCR SPIN, a National Centre of Competence in Research, funded by the Swiss National Science Foundation (grant number 225153).

\clearpage

\bibliography{sn-bibliography}
\end{document}